\input psfig
\documentstyle{l-aa}     
\newcommand{\ROSAT}{{\it ROSAT }}
\newcommand{\RASS}{{\it ROSAT} All-Sky Survey }
\newcommand{\EXSAS}{{\it EXSAS }}
\newcommand{\HEAO}{{\it HEAO-1 }}
\newcommand{\etal}{et~al.\,}

\newcommand{\DEG}{^\circ}
\newcommand{\avrg}[1]{\langle{#1}\rangle}
\newcommand{\com}[1]{}
\newcommand{\chg}[1]{#1}
\begin{document}
\thesaurus{ 02        
(11.01.2   
11.03.1   
12.04.2   
13.25.2   
13.25.3)  
}

\title{A Large-Area Cross-Correlation Study of High Galactic Latitude
          Soft and Hard X-ray Skies}     


\author{Takamitsu Miyaji       \inst{1}
\and   G\"unther Hasinger      \inst{2}
\and   Roland Egger        \inst{1}
\and   Joachim Tr\"umper   \inst{1}
\and   Michael J. Freyberg   \inst{1} }
\offprints{T.\ Miyaji}

\institute{Max-Planck-Institut f\"ur extraterrestrische Physik,
%
D--85740 Garching b.\ M\"unchen, Germany
\and Astrophysikalisches Institut, An der Sternwarte 16, D--14482
Potsdam, Germany}

\date{Received date; accepted date}

\maketitle

\markboth{Miyaji \etal \, Cross-Correlation of Soft and Hard X-ray Skies} 
         {Miyaji \etal \, Cross-Correlation of Soft and Hard X-ray Skies} 


\begin{abstract}
We have made cross-correlation analyses of (2 -- 15 keV) 
\HEAO A2 and 1 keV \ROSAT PSPC All-Sky Survey maps over a 
selected area ($\sim$ 4000 deg$^2$) with high galactic latitude
($b\ga 40\DEG$). We have calculated the correlations for the 
bright \ROSAT sources and residual
background separately with the \HEAO  A2 TOT (2 -- 10 keV)
and HRD (5 -- 15 keV) maps. In any case, the angular dependence of the 
cross--correlation function (CCF) was consistent with expected from the 
\HEAO A2 collimator response function and the smoothing function 
of the \ROSAT 1 keV map. 
The amplitude of the bright \ROSAT source -- A2 CCFs 
are consistent with expectations from model populations 
of AGNs and clusters of galaxies, which emit in both bands. However,
the residual \ROSAT background -- A2 CCFs amplitude at zero degree
are about a factor of three larger than that expected from the model 
populations. 

 Our soft-hard zero-lag and angular CCF results have been compared 
with the 1 keV auto-correlation function (ACF) found by Soltan et al. (1995) 
for the same \ROSAT data. Their significant angular
CCF at a scale of $\theta\la 5\DEG$ is of an unknown origin and it
could represent a new class of cosmic X-ray sources. If
this 1 keV ACF has a hot 
plasma spectrum with $kT\sim 2$ keV, contribution of this
component is  consistent with both our zero-lag 
CCF in excess of the population synthesis model prediction
and the upper-limit to the angular CCF at $\theta\sim 2.5\DEG$.
On the other hand, if this component has a lower temperature
or a steeper spectrum, a major modification  to the
population synthesis model and/or an introduction of new
classes would be needed.    


\end{abstract}
\section{Introduction}

With deep \ROSAT observations of blank fields, up to
$\sim 60$\% of the soft ($\la 2 {\rm keV}$) cosmic X-ray background 
(XRB) has been resolved into individual 
sources (e.g. Hasinger \etal 1993).
A number of groups have worked/are currently working on identification
programs of these \ROSAT sources and we can expect that we will
have the overall picture of the origin of most of the soft ($\sim 1$ keV)
X-ray background including contributing source populations
and their cosmological evolution. 
With ASCA observations of blank fields, the XRB at higher energies 
(2 -- 10 keV) have also been resolved into sources, although
the resolved fraction of the total XRB flux is smaller 
than that of \ROSAT because of the
limited spatial resolution. 
While these observational efforts are
rapidly in progress, the unified theory of AGNs and supporting 
X-ray observations of AGNs for the basic framework of the unified scheme
have provided insights to the origin of the bulk
properties of the XRB.  In particular,
intrinsic photoelectric absorption observed in the 
X-ray spectra of Seyfert 2 galaxies (e.g. Awaki \etal 1991; 
Mulchaey \etal 1992), consistent with the 
view of the unified scheme that a Seyfert 2 galaxy is a Seyfert 1 
seen at an angle where a torus surrounding the central engine 
blocks the line of sight. A wide range of absorption column
densities have been observed (Schartel et al. 1995) for 
a hard X-ray flux-limited sample of AGNs by Piccinotti et al.
(1982). Models of the XRB with evolving populations of unabsorbed and 
internally-absorbed AGNs with various absorption column 
densities, based on the unified scheme of AGNs have been 
successfully made consistent with many observational 
properties of the XRB (Madau \etal 1994; Comastri \etal 1995).
In particular, they explained the global spectral shape
of the XRB, especially  the characteristic break at 30 keV
and source counts in soft and hard X-rays. While these models
are basically successful, there is still room for major modifications,
given current observations. A recent {\it ASCA} observation of NGC 3628 by 
Yaqoob \etal (1995) showed that the X-ray ($E\la$ 10 keV) spectrum 
of this spiral galaxy, which
had previously been classified as a starburst, is rather flat. 
Based on this observation, they argued that the contribution of 
low-luminosity extragalactic X-ray sources, like NGC 3628, to the 
$E \ga 2$ keV XRB could be significant. Such a low X-ray luminosity
population has been deliberately neglected in the Madau \etal (1994) and 
Comastri \etal (1995) models. Also
these models  assumed a simple version of the AGN unified scheme 
where the ratio of the numbers of  absorbed and unabsorbed objects 
does not depend on their intrinsic
X-ray luminosity. However, Barcons \etal (1995) argued that this
is not likely to be the case but rather the fraction of the 
unabsorbed AGNs (type 1) should increase with the intrinsic luminosity.  

 Besides detailed studies of the nature of individual
X-ray sources which can contribute to the XRB, statistical properties 
of the spatial fluctuation of the unresolved background provide us with
other perspectives of cosmic X-ray sources. Fluctuations of the unresolved XRB
have been used to infer the $\log N - \log S$ relation
below the source detection flux-limit (e.g. Shafer 1983;
Hasinger \etal 1993; Barcons et al. 1994). The angular auto-correlation
function is often used
to describe spatial structure of the XRB. Many authors 
discussed the constraints on the combination
of the X-ray volume emissivity and clustering properties of the
X-ray sources from the ACF (or upper limits of ACF) (e.g. Danese \etal 1993;
Carrera \etal 1991; Soltan \& Hasinger 1994). While the XRB ACF traces
the properties of all the X-ray sources along the line of sight and thus
the interpretation is highly model dependent,
cross-correlating catalogs of known sources with the XRB gives
more concrete information on the XRB constituents.  
In particular, cross-correlation functions (CCF) of unresolved XRB spatial 
fluctuations with galaxy catalogs
give information on the contribution of these galaxies and
objects clustered with these sources to the XRB (Jahoda \etal 1991, 1992; 
Lahav \etal 1993; Miyaji \etal 1994; Roche \etal 1994; Barcons \etal 1995). 

 Auto and cross-correlation function analyses over
a large region of the sky occasionally discover new components
of cosmic X-ray emission. The auto-correlation of the \HEAO A2
hard X-ray map nearly over the whole sky at high galactic latitudes
revealed a weak large-scale component at $\theta \la 40\DEG$, 
which perhaps is associated
with structures in the supergalactic plane (Jahoda 1993).  
 Using the \ROSAT All-Sky Survey (RASS) data, Soltan \etal (1995)
found an extragalactic extended component in the ACF of 
at 1 keV at a scale of 5 degrees. They also found an angular
correlation at 5 -- 10 degree scales in the CCF between
the 1 keV XRB and the positions of the Abell clusters. They modeled 
the latter component as a 10 Mpc scale diffuse gas surrounding 
clusters of galaxies, which itself had not been recognized in any 
previous observation. They found, however, that clusters
of galaxies, considering this putative circum-cluster 
emission and clustering of clusters, explain only about a 
third of their extended 1 keV ACF component. Thus the main portion
of the extended ACF could  represent still a new population of 
cosmic X-ray emission and investigating their nature 
is only possible by statistical analyses over a large region 
of the sky.

In this paper, we present the results and possible 
interpretations of a cross-correlation analysis
between the soft ($\sim$1 keV) X-ray sky observed with \ROSAT PSPC
All-Sky Survey and the hard (2 -- 15 keV) X-ray
sky observed  with the \HEAO A2 experiment.
The purpose of the correlation between soft-hard X-rays
presented in this paper is two-folded. Zero-lag and angular
cross-correlation functions (CCFs) between surface brightnesses
contain information on X-ray sources common to these two bands.  
Thus this would give additional constraints to the population
synthesis models. Especially, it would give a check to the 
models based on unabsorbed and absorbed AGNs.
A large area correlation study also may give a clue to the 
nature of the extended ACF component that Soltan \etal (1995) found.  
The scope of this paper is as follows: In Sect.\ref{sec:dat}, 
we explain the data used. We explain the CCF calculation and 
present the results
of the calculation in Sect. \ref{sec:ccf}.  Sect.\ref{sec:form}
presents formulations relating observed CCF with models including
individual sources in the population synthesis models and
extended components. In Sect.\ref{sec:mod}, we use the
formulation to compare various models with the observations.  
The purpose of the section is to check the effects of various components
on the observed CCF and we do not intend to construct a complete 
model consistent with all the existing constraints. Such an elaborate 
modeling would be a topic of a future paper. Finally we conclude 
our discussion in Sect. \ref{sec:conc}.          

\section{Data}
\label{sec:dat}
\subsection{The \ROSAT All-Sky Map}
\label{sec:rosdat}

\begin{figure}[htbp]
\psfig{file=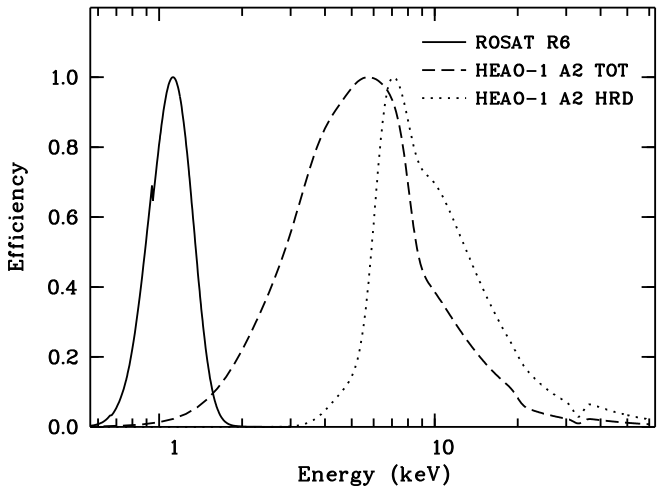,width=\hsize,clip=}
\caption[]{Energy response curves of the \ROSAT R6 and \HEAO A2 TOT/HRD
bands, used in our CCF analysis, are shown as labeled. 
Each curve is normalized at its peak}
\label{fig:resp}
\end{figure}

We have used a clean surface brightness map 
constructed from the \ROSAT (Tr\"umper 1983) PSPC (Pfeffermann et al. 1986)
All-Sky Survey (Snowden \& Schmitt 1990; Voges 1992) in the north galactic
pole region. In making the map, the period of Short-Term Enhancements 
has been excluded and non-cosmic backgrounds (particle background, 
solar scattered X-rays and the Long-Term Enhancements) have been
subtracted as explained in Snowden \etal (1995).  
Soltan \etal (1995) selected a relatively large portion
of the high galactic latitude sky which is free from contamination 
by galactic structures to investigate the extragalactic structures 
in the soft X-ray sky: 
\begin{equation}
70\DEG\, < \, l \, < 250\DEG , \;\; b>40\DEG.
\label{eq:rosreg}
\end{equation}
We also use this area for our correlation with the hard X-ray
sky. We also included a 3.5 degree wide strip surrounding this region
in our analysis, taking the larger \HEAO A2 collimator response function
into account. In this work, we have only used the map in 
the R6 (PI channel range 91 -- 131, corresponding to 0.9 -- 1.3 keV) 
band (Snowden \etal 1993), which is least contaminated by the 
galactic emission, 
scattered solar X-rays, and particle background. The energy response
curve of the R6 band is shown as a solid line in Fig.~\ref{fig:resp}.

The conversion factor between the observed R6 count rate and the 
0.5 -- 2 keV flux before the absorption by the Galaxy 
is 3.24$\times$10$^{-11}$ erg s$^{-1}$cm$^{-2}$/(R6 cts s$^{-1}$)
for a power-law photon index of ($\alpha_{ph}=2.0$) for the
absorption column density of $N_{\rm H}=1.8\times 10^{20}{\rm cm^{-2}}$, which
is the average galactic value for the region. The range of the galactic 
column in the region is $0.6 - 4. \times 10^{20} {\rm cm^{-2}}$ and the
conversion factors in that region differ from the mean value
by $\la 5\%$. Therefore we use the average $N_{\rm H}$ value for the
whole region for the analysis. 

\subsection{The \HEAO A2 Hard X-ray Map}

We have used the all-sky hard X-ray map constructed from
the combination of the HED3 and MED detectors of the \HEAO A2
experiment (Rothschild \etal 1979). The combination of the detectors 
is sensitive in 2--60 keV. Conventionally all-sky
maps from this combination are analyzed in a number of 
standard colors defined as weighted sums of counts from certain
combinations of detector layers and pulse height channels 
(discovery scalers; see Allen \etal 1994). 
In this work, we have used the {\em TOTAL} (indicated by TOT) and 
{\em HARD} (indicated by HRD) band maps. The energy response
curves of the TOT and HRD bands are shown as dashed
and dotted lines respectively in Fig.~\ref{fig:resp}.  
For the cosmic X-ray background and canonical
AGN spectrum, most of the photons come from the 2 -- 10 keV range
for the TOT map and 5 -- 15 keV range for the HRD map. For a power-law
spectrum with photon index $\alpha_{\rm ph}=1.7$, the conversion
factor between the TOT  count rates and the 2--10 keV flux is 
$2.2\times10^{-11}$ erg s$^{-1}$ cm$^{-2}$ (TOT cts s$^{-1}$)$^{-1}$ 
and the conversion
factor between the HRD count rates and the 5 -- 15 keV flux is 
$1.8\times10^{-11}$ erg s$^{-1}$ cm$^{-2}$ (HRD cts s$^{-1}$)$^{-1}$. 
These conversion factors
vary only by $\la 4\%$ for a 40 keV bremsstrahlung spectrum, an $\alpha=0.7$
power-law, the same power-law with neutral gas absorption
up to $N_{\rm H}\la 10^{24} {\rm cm^{-2}}$.

The maps used here have the beam profile (or the collimator 
response function) well represented by (Shafer 1983):
\begin{equation}
B_{A2} = \max(1-\frac{|x|}{3\fdg 0}, 0)\, \max(1-\frac{|y|}{1\fdg 5}, 0)
\label{eq:a2beam}
\end{equation}  
where $y$ represents the coordinate along the scan direction 
of the survey, which is along the great circle of a constant 
ecliptic longitude, and $x$ is the coordinate  
along the direction perpendicular to the scan path. 

\subsection{Excluded Regions}
\label{sec:exc}

Regions around conspicuous sources in the R6 band have been excluded
from the analysis in order to avoid the situation that a few
sources dominate the correlation signal.  
Those are the Coma cluster, Abell 1367, Mkn 421 and 
NGC 4253 (Mkn 766). These are all the sources in the region 
with R6 count rates greater than 0.9 counts per second.
We have excluded 3$\fdg$4 radius regions (corresponding
to the maximum extent of the A2 beam) around Mkn 421 and
NGC 4253. The exclusion radius around the two conspicuous
galaxy clusters (i.e. the Coma cluster and Abell 1367) was 
7$\DEG$.     
 In addition, we have also excluded the 3.4 degree radius regions 
around the \HEAO A2 sources in Piccinotti \etal (1982).  
The flux cutoff of this complete flux-limited catalog is 
1.25 counts s$^{-1}$ in the \HEAO A2 R15 band, which 
is similar to the TOTAL band used here but
a little harder (Allen \etal 1994). 

\section{The Soft-Hard Cross-Correlation Function}
\label{sec:ccf}
\subsection{The CCF calculation}
\label{sec:ccfc}
We have calculated the angular cross-correlation functions
between the \RASS R6 and \HEAO A2 maps in the region
defined in Eq.(\ref{eq:rosreg}). Since the \ROSAT All-Sky map has 
a much higher spatial resolution and denser samplings, we have 
smoothed the {\em ROSAT} data before correlating. 
As the smoothing function, we took the
form represented by Eq. (\ref{eq:a2beam}) but smaller FWHMs
in order to improve the signal-to-noise ratio of the correlation. 
Among a number of sizes we tried for the smoothing function, 
the best results
could be obtained when we chose $1\fdg 5$ and $0\fdg 75$ (FWHMs) 
in the x and y directions respectively.

As a statistical measure to indicate the degree of correlation
between two maps (here we denote the two maps by subscripts ${\rm s}$
and ${\rm h}$, representing {\it soft} and {\it hard} respectively),
we calculate the angular cross-correlation function (CCF):
\begin{equation}
W_{\rm sh}(\theta) = \frac{N_{\rm pair}^{-1}\sum_{ij} 
\left[ (I_{{\rm s}i}-\avrg{I_{\rm s}})\, (I_{{\rm h}j}-\avrg{I_{\rm h}}) 
\right] }
{\avrg{I_{\rm s}}\,\avrg{I_{\rm h}}},
\label{eq:wsh}
\end{equation}
where $I_{\rm s}$ and $I_{\rm h}$ are the \ROSAT and \HEAO A2 intensities
measured at the gridding points which are within the region
defined above.  We have taken pairs along the scan path 
(ecliptic longitude) to calculate the angular correlation in 
order to make use of the narrower collimator response.

The uncertainties of the correlation functions are estimated by
correlating the \ROSAT map with different parts of the sky
in the \HEAO A2 map. We made this by rotating the \ROSAT map 
around the galactic pole and calculating the correlation with the A2 
map in the same way.   
We have also transposed the \ROSAT map
about the galactic equator followed by rotations about
the galactic pole to calculate the correlation with the 
A2 map. Thus we produced a number of different
artificial null samples with equivalent statistical properties.
Regions around Piccinotti \etal sources are excluded as 
we did in the real correlation calculation. We also excluded 
20$\DEG$ and 25$\DEG$ regions around LMC and SMC
respectively. We have rotated the map in steps of  30$\DEG$
on both hemispheres, thus we get 23 null-hypothesis CCFs
between unrelated parts of the sky. These give a fair estimate 
of the uncertainties of the correlation function.

\subsection{CCF of \ROSAT bright sources with the \HEAO A2 Maps}
\label{sec:r6src}

In order to evaluate the contribution to the CCF from bright
\ROSAT sources, we have made a list of bright sources 
by performing a source detection procedure to the
R6 map in the region defined in Eq.(\ref{eq:rosreg}). The source
detection  has been made to the map gridded in $12'\times 12'$ pixels, 
larger than the \ROSAT PSPC point spread function, and thus
is much less sensitive than the source detection procedure 
directly applied to the full-resolution \RASS data. 
A \RASS bright source list using such a 
procedure (Voges 1995) is in progress at the time
of writing this paper.  An analysis using the source list 
will be a topic of a future paper. In this work, we use 
the coarsely rebinned map to obtain a complete R6 flux-limited sample of
bright sources.  As described by Hasinger \etal  (1993), 
the source detection have been made in two steps: the first 
step is to slide a detection cell
over the map and find excess over the region immediately
surrounding the cell (LDETECT). A background map was created by
removing the sources detected in the LDETECT procedure, 
filling the holes by the mean counts surrounding them, and 
making a bi-cubic spline fit to the filled map. \chg{This procedure
gives a more robust measure of the backgound for the source
detection than taking the background from immediate surroundings.} 
Then we slid the detection cell again to the original map to find 
significant excess counts over the background map created in the 
above process (MDETECT). 
 
The source detection have been made using 
procedures within the \EXSAS software package
(Zimmermann et al. 1994).
The size of the detection cell has been  taken to be 3$\times$3 cells 
$(36'\times 36')$ and the source detection threshold is set 
at a significance of  $-\log P_i = 10$, where $P_i$ is the probability
that the random Poisson fluctuation make the excess counts in the cell.
This roughly corresponds to a 4$\sigma$ excess of a Gaussian fluctuation.
\chg{Since the \ROSAT PSPC PSF and  extents of clusters
of galaxies (except for closest ones, which have been excluded
from the analysis (Sect. \ref{sec:exc}) are much smaller than 
the size of the detection cell, 
the source list from the MDETECT procedure is expected to be free 
from significant biases.}

In the least sensitive areas, the detection limit was 0.08 R6 
cts s$^{-1}$. \chg{If we assume a Seyfert 1-like broken power-law 
spectrum with $\alpha_{ph}=2.3$ for $E\le 1.5$ keV, 
$\alpha_{ph}=1.7$ for $E>1.5$ keV, and the average galactic 
$N_{\rm H}$ of the region
($=1.8\times10^{20}\,{\rm cm}^{-2}$), 0.08 R6 cts s$^{-1}$ corresponds 
to a 0.5 -- 2 keV flux of 2.7$\times$10$^{-12}$ ${\rm erg\,cm^{-2}\,s^{-1}}$ 
and 2 -- 10 keV flux of 3.8$\times$10$^{-12}$ ${\rm erg\,cm^{-2}\,s^{-1}}$.}
 
By this procedure, we have obtained a complete flux-limited
($>0.08 {\rm R6\, cts\,s^{-1}}$) 
sample of 183 sources above this limit over the 4575 deg$^2$ area of 
the sky (0.04 sources deg$^{-2}$). \chg{The  
$\log\,N$ -- $\log S$ function of the detected sources
has a Euclidean slope. 
The number density corresponds to that of the {\it Einstein} 
Medium Sensitivity Survey (EMSS)  (Gioia \etal 1990) above 
4.1$\times$10$^{-12}$ ${\rm erg\,cm^{-2}\,s^{-1}}$ in 0.3 -- 3.5 keV.
They are consistent if  the 0.3 -- 3.5 keV flux versus R6 count rate 
conversion factor is $\sim 5\times 10^{-11}$ ${\rm erg\,s^{-1} cm^{-2}\, 
(R6\,cts\,s^{-1})^{-1}}$. For a power-law spectrum with $\alpha_{\rm ph}
=2.0$,
this factor is  $\sim 5.7\times 10^{-11}$ ${\rm erg\,s^{-1}\, cm^{-2}
\, (R6\,cts\,s^{-1})^{-1}}$ and for a $kT$ = 2 keV 
Raymond-Smith spectrum (heavy element abundance=0.4 cosmic), this
number is $\sim 4.4\times 10^{-11}$ ${\rm erg\,s^{-1}\,cm^{-2}\, 
(R6\,cts\,s^{-1})^{-1}}$. Thus our bright source counts are 
consistent with the EMSS counts considering a possible range of
the source spectra.}    

We have calculated the CCF between the detected sources
(R6 count rate weighted) and the \HEAO A2 maps in the same
way as Sect. \ref{sec:ccfc}. The errors of these CCFs have 
been calculated also in the same manner.

Using the detected sources, we have also constructed a
{\em source-removed} \ROSAT map. This has been made by replacing the
3$\times$3 image pixel regions around the sources with R6 count rate
greater than 0.08 cts s$^{-1}$ by the pixel values
in the same regions of the background map produced for the MDETECT 
procedure (see above).  We have also calculated the CCF between 
the source-removed map and the \HEAO A2 maps.\chg{The
source-removed map still contains sources below the limit
and we expect that such fainter sources still 
contribute to the correlation.}  

\subsection{Results of the CCF Calculations}

\begin{table*}   
\caption[ ]{The Zero-lag Correlation Results}
\begin{flushleft}
\begin{tabular}{llllll}
\noalign{\smallskip}\hline\noalign{\smallskip}
Id. & R6 cr. range & \HEAO  & $\avrg{I_{\rm s}}$ & $\avrg{I_{\rm h}}$ & $W_{\rm sh}(0)$ \\
& [cts s$^{-1}$] & A2 Band & \multicolumn{2}{c}{[cts s$^{-1}$ deg$^{-2}$]}
&  \\
\noalign{\smallskip}\hline\noalign{\smallskip}
T1  & 0 -- 0.9   & TOT & $2.9\times 10^{-1}$ & $7.2\times 10^{-1}$ 
& $(3.2\pm 0.4)\times 10^{-3}$ \\
T2  & 0 -- 0.08  & TOT & $2.9\times 10^{-1}$ & $7.2\times 10^{-1}$
& $(1.7\pm 0.3)\times 10^{-3}$ \\
T3  & 0.08 -- 0.9 & TOT & $6.1\times 10^{-3\,\rm a}$ & $7.2\times 10^{-1}$  
& $(6.6\pm 1.0)\times 10^{-2}$ \\
T4  & 0.2 -- 0.9 & TOT & $2.6\times 10^{-3\,\rm a}$ & $7.2\times 10^{-1}$ 
& $(8.3\pm 1.7)\times 10^{-2}$ \\

H1  & 0 -- 0.9   & HRD & $2.9\times 10^{-1}$ & $8.6\times 10^{-1}$ 
& $(2.4\pm 0.4)\times 10^{-3}$ \\
H2  & 0 -- 0.08  & HRD & $2.9\times 10^{-1}$ & $8.6\times 10^{-1}$
& $(1.0\pm 0.3)\times 10^{-3}$ \\
H3  & 0.08 -- 0.9 & HRD & $6.1\times 10^{-3\, \rm a}$ & $8.6\times 10^{-1}$  
& $(5.4\pm 1.6)\times 10^{-2}$ \\
H4  & 0.2 -- 0.9 & HRD & $2.6\times 10^{-3\,\rm a}$ & $8.6\times 10^{-1}$ 
& $(6.6\pm 2.6)\times 10^{-2}$ \\
\noalign{\smallskip}\hline
\end{tabular}
\label{tab:ccf0}
\begin{list}{}{}
\item[$^{\rm a}$]Intensity from sources in the quoted flux range 
\end{list}
\end{flushleft}
\end{table*}

Significant zero-lag correlation signals have been detected in most of
the cases we have calculated ($\theta=0$ case of Eq. [\ref{eq:wsh}]). 
Table \ref{tab:ccf0} summarizes 
the zero-lag correlation results for the calculations we made for 
various cases.
The second column of Table \ref{tab:ccf0} is the R6
count rate range of the sources correlated with the A2 map. A
zero value of the lower bound in this column means that 
the R6 surface brightness map has been used and a non-zero value 
represents that the source list in the quoted flux range explained in  
Sect. \ref{sec:r6src} has been used.  

\begin{figure}[htbp]
\psfig{file=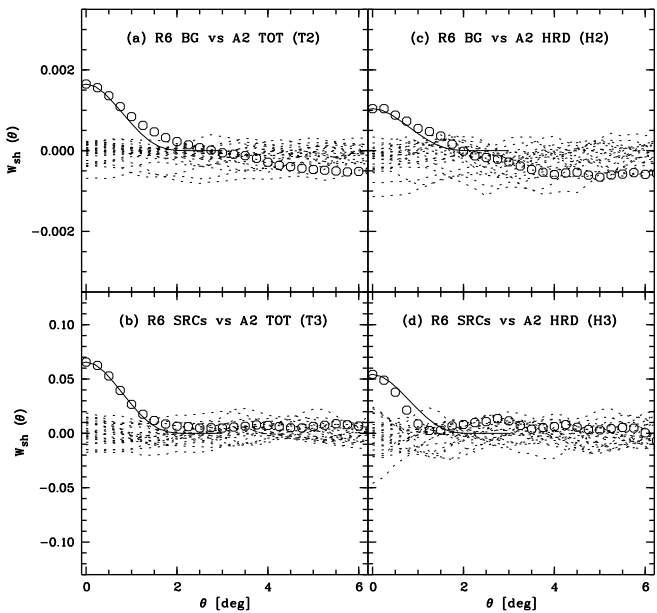,width=\hsize,clip=}
\caption[]{The angular CCF for selected cases as labeled. 
The open octagons show the
observed correlation and the dotted lines show the 23 uncorrelated
cases. The solid line is the angular dependence of the Poisson effect
caused by the A2 collimator response function and smoothing}
\label{fig:wth}
\end{figure}

The angular correlation functions for selected cases have been
plotted in Fig. \ref{fig:wth} along with the angular correlation
functions between the rotated R6 maps with the A2 map to show the
dispersion of the CCF when no real sky-correlation exists. 
The label in each figure corresponds to the correlation
ID in Table \ref{tab:ccf0}. Also the angular
dependence of the CCF in case of the purely Poissonian case 
(see Sect. \ref{sec:form}) due to the
A2 collimator reponse and the \RASS map smoothing, normalized at
$\theta=0$, has been plotted on each plot. In any case, we 
did not find significant extended component in the angular CCF beyond 
expected from the Poisson effect.  

\section{Basic Formulations}
\label{sec:form}

The CCF depends on  Poisson and  clustering effects. 
The Poisson effect is caused by individual 
sources which emit both soft and hard X-rays and the
angular dependence of the correlation coefficient simply reflects
the overlap of the finite-sized beams of soft and hard X-ray observations
and smoothings. The clustering  effect reflects the real sky structure of 
the X-ray sources common to (or at least correlated with) both 
soft and hard X-rays. This structure may be due to clustering
of sources or a diffuse X-ray component which extends to a scale
comparable or larger than the size of the beam. If one is concerned
with real extended structure, the distinction between Poisson
and clustering effects is somewhat ambiguous. In our case,
the extent of the X-ray emission from an individual galaxy cluster
is much smaller than the beam size of the \HEAO A2 experiment
and thus we treat them  in the Poisson term. On the other hand, 
we treat the effect of possible diffuse components with scales of 
several degrees in the clustering effect. 

The observed angular correlation function is then, expressed by the
sum of two terms (cf. Lahav et al. 1993; Miyaji 1994; Miyaji \etal 1994):
\begin{equation}
W_{\rm sh}({\theta})\avrg{{\cal I}_{\rm s}}\avrg{{\cal I}_{\rm h}} = 
\eta_{\rm p} (\theta) + 
\eta_{\rm c} (\theta),
\label{eq:wpc}
\end{equation}
where ${\cal I}_{\rm s,h}$ represent the intensities of soft and hard X-rays
per {\em beam} respectively:
\begin{equation}
{\cal I}_{\rm s,h} \equiv \int \avrg{I_{\rm s,h}}\,B_{\rm s,h}
(\hat{\Omega})\,
{\rm d}\Omega
\equiv \avrg{I_{\rm s,h}}\Omega_{\rm Bs,h},
\end{equation}
where $\hat{\Omega}$ is the unit vector towards the solid angle
element ${\rm d}\Omega$. 

\chg{As we will see in Sect. \ref{sec:mod}, the correlation
is dominated by local sources and cosmological evolution is
not important in our case, because of the large beam of the
\HEAO A2 experiment. However, we will develop a full cosmological
formulation below for completeness.}

\subsection{Poisson Effect from Common Sources}

We can express the Poisson term of the 
cross-correlation function as:
\begin{eqnarray}
\eta_p (\theta) = \int \int S_{\rm s}\,S_{\rm h}\, 
N_{\rm sh}(S_{\rm s},S_{\rm h})\,{\rm d}S_{\rm s} {\rm d}S_{\rm h}\, \times
\nonumber \\
\avrg{\int B_{\rm s}(\hat{\Omega})\,
B_{\rm h}(\hat{\Omega}-\vec{\theta})\,{\rm d}\Omega},
\label{eq:etap}
\end{eqnarray}  
where $N_{\rm sh}(S_{\rm s},S_{\rm h})\,
{\rm d}S_{\rm s},{\rm d}S_{\rm h}$ is the {\em bivariate} $N(S)$ function 
defined as the number of sources per solid angle whose soft and hard X-ray 
fluxes fall in the ranges $[S_{\rm s}-0.5\,{\rm d}
S_{\rm s},S_{\rm s}+0.5\,{\rm d}S_{\rm s}]$ and 
$[S_{\rm h}-0.5\,{\rm d}S_{\rm h},S_{\rm h}+0.5\,{\rm d}S_{\rm h}]$ 
respectively. Also $\vec{\theta}$ is the
offset vector between hard and soft X-ray measurements. 
The flux integrations (over $S_{\rm s}$ or $S_{\rm h}$) 
are over source fluxes 
included in the corresponding maps. When we use a surface brightness map 
where bright sources above a certain limit are excluded, 
this limit defines the upper bound of the integration and the lower
bound is zero. When we use a complete flux-limited source list, this
flux limit defines the lower bound.

The double integral over the soft and hard fluxes 
in Eq. (\ref{eq:etap}) can be calculated for given population 
synthesis models. Let us suppose we have evolving
populations of objects with the bivariate soft-hard luminosity
function (for the $i$-th population) per comoving volume:
$\hat{\Phi}_i(L_{\rm s},L_{\rm h},z)\,{\rm d}L_{\rm s}\,{\rm d}L_{\rm h}$,
where each population is assumed to have a universal spectral shape
for simplicity. Then,
\begin{eqnarray}
\int S_{\rm s}\,S_{\rm h}\,N(S_{\rm s},S_{\rm h})\,{\rm d}S_{\rm s}\,
{\rm d}S_{\rm h}\,=\, \frac{c}{16\pi^2 H_0} 
\int_0^{z_{\max}}\times  \nonumber \\
\frac{1}{D_{\rm L}(z)^2 (1+z)^3(1+\Omega_0z)^\frac{1}{2}}\,
\sum_{i=pop}
\left[ K_{{\rm s}i}(z)K_{{\rm h}i}(z) \times \right. \nonumber \\
\, \int_{L_{\rm s}^{\min}}^{L_{\rm s}^{\max}} 
\int_{L_{\rm h}^{\min}}^{L_{\rm h}^{\max}} L_{\rm s} L_{\rm h} 
\hat{\Phi}_i(L_{\rm s},L_{\rm h},z){\rm d}L_{\rm h}\,
{\rm d}L_{\rm s} \left. \right]\,dz,
\end{eqnarray}
where  $D_{\rm L}(z)$ is the luminosity distance as a function of redshift 
and $K_{{\rm s(h)}i}$ is the soft (hard) band K-correction for the $i$-th
population. The luminosity integrations are 
over luminosities where the observed flux falls in the range 
of source fluxes included in the corresponding correlating maps. 
Thus the bounds of the luminosity integrations depend on 
redshift and the spectrum of the population. Now let the 
soft and hard X-ray luminosities of
each population be related by $L_{\rm s}=a_iL_{\rm h}$. 
Then the double integral over hard and soft X-ray luminosities 
can be reduced to:
\begin{eqnarray}
\int_{L_{s}^{\min}}^{L_{s}^{\max}} \int_{L_{\rm h}^{\min}}^{L_{h}^{\max}}
L_{\rm s} L_{\rm h} 
\hat{\Phi}_i(L_{\rm s},L_{\rm h},z){\rm d}L_{\rm h}\,{\rm d}L_{\rm s} \nonumber \\
= \int_{\max (L_{\rm s}^{\min}\,\, , a_iL_{\rm s}^{\min})}
^{\min (L_{\rm s}^{\max}\,\, ,a_iL_{\rm h}^{\max})} 
a_i\, L_{\rm s}^2\, \hat{\Phi}_{\rm s}(L_{\rm s},z)\,{\rm d}L_{\rm s},
\end{eqnarray}  
where $\hat{\Phi_{\rm s}(L_{\rm s},z)}$ is the comoving soft X-ray luminosity
function, which we can find in literature based on \ROSAT surveys (e.g.
Boyle et al. 1993, 1994). 
In the case of the pure-luminosity evolution 
of the power-law form, $L(z)=L(0)(1+z)^p$, the evolving 
luminosity function can be expressed by the
present epoch one:
\begin{equation}
\hat{\Phi}_{{\rm s}i}(L_{\rm s},z) = \hat{\Phi}_{\rm{s}i}([1+z]^{-p}
L_{\rm s},0)(1+z)^{-p}. 
\end{equation}

\subsection{Extended Component}

The underlying angular CCF of the soft and hard X-ray skies
\begin{equation}
w_{\rm sh}(\theta)\equiv \frac{\avrg{I_{\rm s}(\hat{\Omega})\,I_{\rm h}
(\hat{\Omega}+\vec{\theta})}}{\avrg{I_{\rm s}}\avrg{I_{\rm h}}}-1, 
\hspace{2em} 
\theta>0
\end{equation}
reflects the extended structure common to soft and hard X-rays and
clustering of X-ray sources. 

The {\em observed} angular correlation function
($W_{sh}(\theta)$) is contributed by the underlying angular
cross-correlation function $w_{\rm sh}(\theta)$ of the real sky. The
term $\eta_{\rm c}$ in Eq. \ref{eq:wpc} reflects this component: 
\begin{eqnarray}
\eta_{\rm c} (\theta) =  \avrg{I_{\rm s}} \avrg{I_{\rm h}} \times \hfil 
\nonumber \\
\int \int w_{\rm sh}(\theta ')
B_{\rm s}(\hat{\Omega}_1)\,B_{\rm h}(\hat{\Omega}_2 - \vec{\theta})\, 
{\rm d}\Omega_1\,{\rm d}\Omega_2,
\label{eq:etac}
\end{eqnarray}
where $ \theta' = |\hat{\Omega}_1 - \hat{\Omega}_2 + \vec{\theta}|$.
In effect, the underlying angular correlation function is
smoothed with the convolution of soft and hard beams.
At large separations ($\theta$ is much larger than the 
scale size of the beam), $W_{\rm sh}(\theta) \approx  w_{\rm sh}(\theta)$
and the Poisson term is zero.

The underlying $w_{\rm sh}$ can be expressed by the cosmologically evolving
soft and hard volume emissivities and the spatial correlation function
of soft and hard X-ray sources. The expression is quite parallel
to the case of ACF found in literature
(e.g. Danese \etal 1993; Soltan \& Hasinger 1994). 
Let us consider populations of soft and hard X-ray emitting sources
with redshift dependent comoving volume emissivities of 
$\hat{\rho}_{\rm s}(z)$ and $\hat{\rho}_{\rm h}(z)$.
Let the spatial correlation 
function between these populations be $\xi_{\rm sh}(r,z)$. 
Then, in the small-angle long-distance approximation,
\begin{eqnarray}
w_{\rm sh}(\theta)=\frac{c}{16\pi^2H_0}\int\hat{\rho}_{\rm s}(z)K_{\rm s}(z)
\hat{\rho}_{\rm h}(z)K_{\rm h}(z) \times \nonumber \\
\int \xi_{\rm sh}(\sqrt{(D_A \theta)^2+x^2},z)\,{\rm d}x
(1+z)^{-4}(1+\Omega_0z)^{\frac{1}{2}}{\rm d}z,
\label{eq:wshc}
\end{eqnarray}
where $K_{\rm s}(z)$ and $K_{\rm h}(z)$ are the K-corrections for the
soft and hard X-ray bands respectively and $D_A(z)$ 
$(=D_L(z)(1+z)^{-2})$ is the angular distance.

\section{Comparison of Models with Observations}
\label{sec:mod}

\subsection{AGN and Cluster Contributions}
\label{sec:mod_p1}

We have calculated the expected contributions from two major
classes of extragalactic X-ray sources, i.e. AGNs and
clusters of galaxies, to the observed zero-lag correlation 
strengths using the population synthesis technique. For the following
models, the standard cosmology with $\Lambda=0$ and
$q_0=0$ has been used. We also denote the Hubble
constant by $H_0=50h_{50} {\rm km\,s^{-1}\,Mpc^{-1}}$.
Note that the expected correlations from the models
do not depend on the value of the Hubble constant.

As the first model (Model P1, where P represents the
Poisson effect) we have constructed an AGN population
synthesis model following the recipe by Comastri \etal (1995)
for their baseline model.
Contributions from X-ray clusters of galaxies
have been added. In adding the cluster contribution, we have used
the analytical forms of the 2 -- 10 keV X-ray luminosity
function (XLF) by Edge \etal (1990). We have used their 
volume-limited expression
for $z<0.1$ and flux-limited expression for $0.1\leq z < 0.17$, taking
into account the deficit of most luminous clusters at
higher redshifts they observed. 
In the $0.17\leq z <0.6$, we have used the power-law
forms of the XLF given by Henry \etal (1992) in their three redshift 
bins. They showed the steepening of the XLF for higher redshift bins.
Thus their power-law form would predict too large an XLF at low
X-ray luminosities (below their sample limit). 
Thus we take the smaller (in comoving coordinates) 
of the Edge et al. analytical form and  the Henry et al. power-law
form. We have taken the 2--10 keV 
X-ray luminosity range of 
$0.01\leq L_{\rm x44} \leq 120$, where 
$L_{\rm x44}$ represents X-ray luminosity measured
in units of $10^{44}h_{50}^{-2}{\rm erg\,s^{-1}}$. 
The  clusters have been assumed to have X-ray
spectra represented by the Raymond \& Smith plasma with 
a heavy element abundance of 0.4. The temperatures of a cluster 
have been chosen from $kT=$ 1, 1.5, 2.5, 4, 7, 10 keV,
according to its luminosity between the minimum and the
maximum values given above. The dividing luminosities are
$L_{\rm x44}=$ 0.02, 0.1, 0.5, 3.0, and 15.0, which roughly corresponds
to the X-ray luminosity versus  temperature correlation (David \etal 1993).
There is a significant scatter in the luminosity -- temperature
correlation but this has not been taken into account in the model. However,
the results are insensitive to the detailed assumption on this 
$L_{\rm x}$ -- $T$ correlation. 
For example, assuming a single temperature of $kT = 4$ keV for all
clusters changed the result by 10\% for T2, which 
is somewhat smaller than the statistical error of the observation. 
The expected correlation from this model is compared with data
in Table \ref{tab:mcomp}. We compare the values of 
$W_{\rm sh}(0)\avrg{I_{\rm s}}\avrg{I_{\rm h}}$ because this quantity 
is proportional to the contribution from each of the R6 sources at a certain 
flux-range and source-removed background.   

\begin{table}   
\caption[ ]{Comparison of correlation results with models}
\begin{flushleft}
\begin{tabular}{lllllll}
\noalign{\smallskip}\hline\noalign{\smallskip}
& \multicolumn{5}{c}{$W_{\rm sh}(0)\avrg{I_{\rm s}}\avrg{I_{\rm h}}$}\\
Id.  & \multicolumn{5}{c}{[$10^{-4}$ R6 cts s$^{-1}$ A2 cts s$^{-1}$]} \\ 
\cline{2-7} 
& Observed & \multicolumn{5}{c}{Models$^{\rm a}$} \\ \cline{3-7}
&          & P1 & P2 & P3 & E1 & E2 \\
\noalign{\smallskip}\hline\noalign{\smallskip}
T2 & $3.5\pm0.6$ & 0.9 & 1.3 & 0.9      & 3.1 & 1.9  \\
T3 & $2.9\pm0.4$ & 2.5 & 3.6 & 2.5      & \ldots & \ldots \\
\\
H2 & $2.5\pm0.7$ & 0.7 & 1.1 & 0.8      & 0.8 & 0.2  \\
H3 & $2.8\pm0.8$ & 1.7 & 2.5 & 1.7      & \ldots & \ldots  \\
\noalign{\smallskip}\hline
\end{tabular}
\label{tab:mcomp}
\begin{list}{}{}
\item[$^{\rm a}$] See Sect. \ref{sec:mod_p1} for the explanation
of model P1, Sect. \ref{sec:mod_p23} for P2 \& P3, and  Sect. \ref{sec:mod_e}
for E1 \& E2 
\end{list}
\end{flushleft}
\end{table}

\begin{figure*}[htbp]
\psfig{file=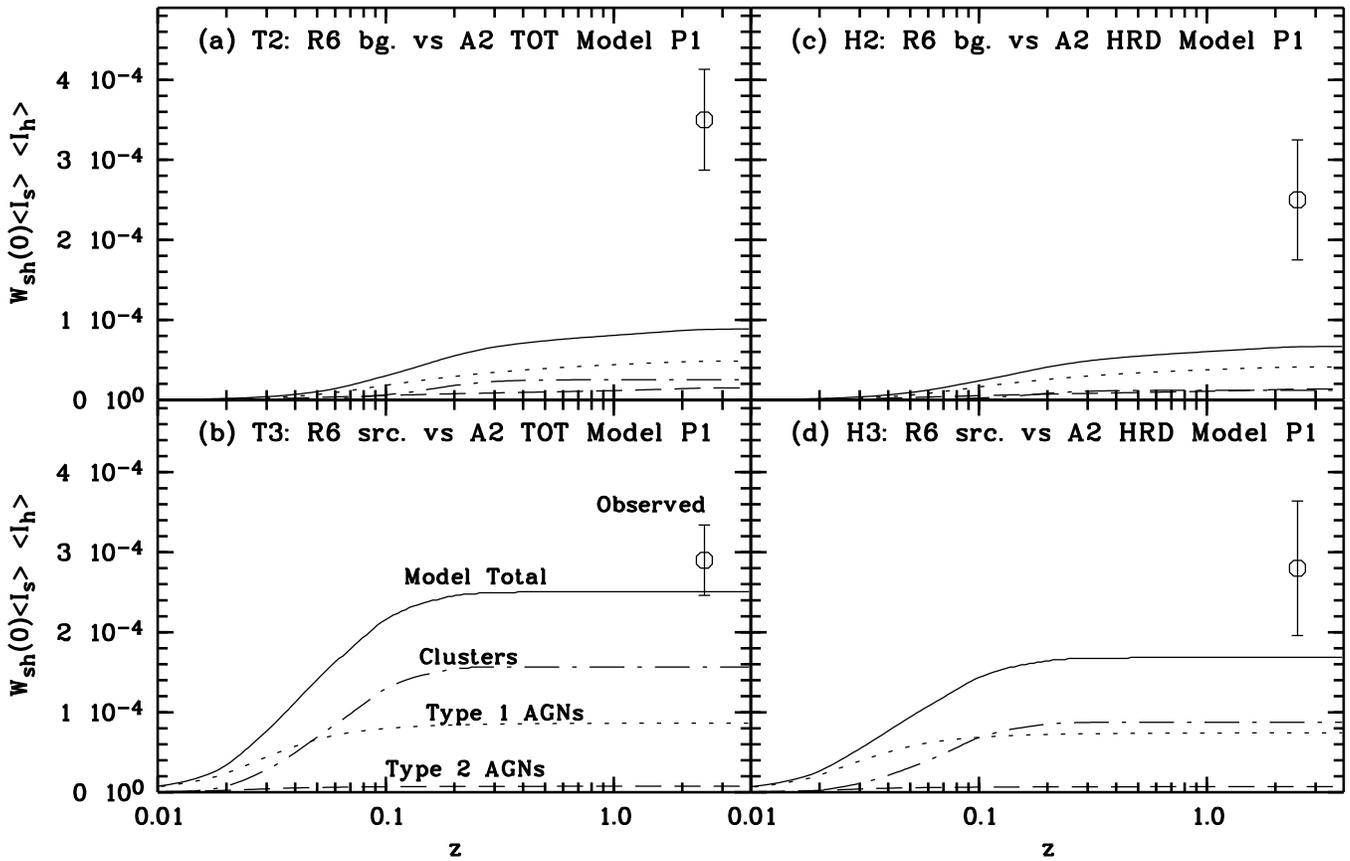,width=\hsize,clip=}
\caption[]{The cumulative contributions of clusters, type 1 and 2 AGNs,
and the total of these for model P1 to the observed CCFs  
are plotted. Each component is drawn  in a different
line style as labeled in the lower left panel. The observed values
with 1$\sigma$ error bars are also shown. The horizontal positioning
of the observed data does not have any significance but placed there
only for the convenience of the display.}
\label{fig:p1_z}
\end{figure*}

Figure \ref{fig:p1_z} shows the cumulative 
contributions of clusters, type 1 AGNs and type 2 AGNs in 
model P1 to the $W_{\rm sh}(0)\avrg{I_{\rm s}}\avrg{I_{\rm h}}$ values for 
selected correlations.  
The observed value with $\pm 1\sigma$ range is shown for each 
correlation. 
As we see in Table \ref{tab:mcomp} and Fig. \ref{fig:p1_z}, model P1 
is consistent within 2$\sigma$
with the observed correlations between A2 maps and bright 
R6 sources (T3 and H3). We also see that these correlations
are dominated by clusters of galaxies. On the other hand, 
the observed correlations between the 
bright source {\em removed} R6 map and the A2 maps are about a 
factor of 3 -- 4 larger than the model predictions. \chg{In any case,
the correlation is dominated by sources in the local universe
($z \la 0.2$) and the expected correlation does not depend
on the cosmological evolution significantly.}  
The contribution of galactic stars, effects 
of minor modifications to the model, and the effects of possible 
extended components are discussed in the followings.  

\subsection{Contribution of Galactic Stars}

Among the 835 serendipitously found X-ray sources in 780 deg$^2$ of
high galactic latitude sky in the {\it Einstein} Extended Medium 
Sensitivity Survey (EMSS), about 
25\% are indentified with galactic stars (Stocke et al. 1991). 
The identification was 96\% complete and the flux limits of 
the survey range between 0.5 -- 34 
$\times 10^{-13} {\rm erg\,s^{-1}cm^{-2}}$ 
in the 0.3--3.5 keV band. When converted to the \ROSAT R6 band, this 
corresponds to a flux range just below the detection limit in our
source detection discussed in Sect. \ref{sec:r6src}.
Based on the flux-number ($\log\,N$ -- $\log\,S$) relation for
the EMSS sample of stars, we have estimated the  contribution of 
Galactic stars to the present correlation
signals. Assuming a $kT=2$ keV Raymond \& Smith plasma spectrum 
with cosmic abundances, which is 
typical of a variable hard component of stellar coronal
emission, the estimated contribution to the R6 background 
versus A2 TOT correlation (T2) is 
$\approx 3\%$ and R6 source versus A2 TOT correlation (T3) is 
$\approx 10\%$ of the observed. Thus, the contributions of
galactic stars to the observed cross-correlations are
within the statistical errors and not significant.     

\subsection{Unabsorbed Soft Component associated with AGNs}
\label{sec:mod_p23}

In our model P1, we used the AGN population synthesis model
by Comastri \etal (1995). We here consider two minor modifications 
to this model.  If a significant fraction
of AGNs are embedded in groups of galaxies with X-ray emitting 
gas with $kT\sim$1 keV, this could contribute to the soft-hard
correlation. In many type 2 AGNs, there is an unabsorbed X-ray emitting 
component, which is 
believed to be electron-scattered X-rays from the nucleus.
Comastri et al. (1995) \etal deliberately neglected this
component because  its luminosity is much
weaker than the soft X-ray luminosity in type 1 AGNs with
a comparable optical luminosity. We see the impact of these
effects to our correlation. 

In model P2,  each AGN has 
an unabsorbed Raymond \& Smith plasma component of $kT=$ 1 keV 
with a 0.5 - 2 keV luminosity of 
$2\times10^{42}h_{50}^{-2}{\rm erg\,s^{-1}}$.
This is a typical value for most luminous groups of galaxies
found in the \ROSAT North Ecliptic Pole Survey (Henry \etal 1995).
The spatial number density of groups of galaxies at this luminosity 
is several times $10^{-5} h_{50}^3 Mpc^{-3}$, which is comparable
to the number density of Seyfert galaxies and about two orders of magnitude
smaller than the number density of bright spiral galaxies. Thus,  
it is unlikely that every AGN is associated with a group of galaxies
at this luminosity. The purpose of this model is not to be realistic 
but to examine the effect of the additional component. 
Table \ref{tab:mcomp} shows that adding this component has little effect
on the correlations T2 and H2, thus fails to explain the excess
correlation even in this extreme case. 

As another model, P3, we put a soft component to AGNs with 
fluxes proportional to the intrinsic (i.e. unabsorbed)
luminosity. The underlying idea of this model, in terms
of the unified scheme of AGNs, is to add an unbasorbed   
X-ray component in Seyfert 2, which is generally considered
as electron-scattered nuclear emission. Like Madau \etal (1994)
considered in their model, we assume that 2\% of the nuclear 
emission in type 2 AGNs are from the scattered component
and can reach us without being absorbed. As we see in 
Table \ref{tab:mcomp}, this component is not sufficient to explain the
observed excess correlation over model P1. 

\subsection{Low-Luminosity Flat X-ray Spectrum Galaxies}

The Comastri et al. (1995) model deliberately neglected the contribution
of low X-ray luminosity objects ($L_{\rm x}\la 10^{42} h_{50}^{-2}
{\rm erg\,s^{-1}}$) to the hard X-rays ($E\ga 2$ keV), while
they left some room for these objects to contribute to soft X-rays
($E\la 2$ keV). However, Yaqoob et al. (1995) recently reported
that NGC 3628, a low X-ray luminosity object, which had usually
been classified as 
a starburst, has a rather flat X-ray spectrum (a photon 
index of $\alpha_{\rm ph}=1.2$). 
Based on this observation, they argued that 
a class of sources represented by this object could contribute 
significantly to the XRB.  Here we estimate the contribution of the 
possible low luminosity population to our correlations. For an 
estimation, we consider low-luminosity X-ray galaxies of 
the 2 -- 10 keV luminosity range of 
$10^{40}-10^{42}h_{50}^{-2}{\rm erg\,s^{-1}}$ at the present epoch.
We assumed a power-law differential lumninosity function 
(${\rm d}\Phi(L_{\rm x})/{\rm d}L_{\rm x} \propto L_{\rm x}^{-\gamma}$) 
with $\gamma=1$ and
total 2 -- 10 keV present-epoch volume emissivity of 
$2\times10^{38}h_{50}{\rm erg\,s^{-1}\,Mpc^{-3}}$. This volume
emissivity is an upper limit from the low-luminosity source
contribution derived by Miyaji et al. (1994) based on the XRB-
{\it IRAS} galaxy cross-correlation and has been considered
in the Yaqoob et al. baseline model. For this population, we
have assumed a power-law spectrum with $\alpha_{\rm ph}=1.2$ 
and simple luminosity evolution of $\propto (1+z)^{3.2}$
up to $z=2$.  
The contribution of this population to our
correlation calculated using above assumptions is only
$\sim 1\%$ for either T2 or T3. Note that, in this model, 
most contribution to the correlation signal comes from around 
$z\sim 0.1$, unlike the contribution to the total
surface brightness, which is mostly contributed from the high-redshift
region given this evolution law. Thus this estimate is insensitive to
the detailed behavior of the cosmological evolution. The above estimation 
shows that the low X-ray luminosity galaxies cannot explain the
observed excess correlation in T2 or H2. 
 
\subsection{Extended Emission Component} 
\label{sec:mod_e}

We consider here the effect of the clustering term (or extended
components) to the observed $W_{\rm sh}(0)$. In the extragalactic
component of the R6 (1 keV) map, Soltan \etal (1995) found a
significant extended component in the auto-correlation 
function (ACF) at a scale of $\theta\la 5\DEG$.
They argued that the ACF signal they have
found should be dominated by
real extended structures rather than clustering of sources,
based on comparison with the ACF in Soltan and Hasinger 
(1994) at smaller scales. They speculated that this 
component could be due to the clustering of groups of galaxies.
However, information on statistical properties of groups 
of galaxies was very scarce and they did not make further 
arguments on this possibility.


 The expected R6 ACF at $\sim 1\DEG$ -- $3\DEG$ from the 
clusterings of AGN and clusters used in model P1 using Eq. \ref{eq:wshc}
(in this case, two bands in the expression are the same)
are two orders of magnitude smaller than the actual values
observed by Soltan et al. (1995). We also checked if the
AGN and cluster clusterings can explain the excess correlation
in our zero-lag CCFs (T2 and H2). The expected clustering
effects to the \ROSAT-- HEAO-1 zero-lag CCFs (because of the
beam smearing, the zero-lag values are mainly due to the 
clustering at a scale of 1$\DEG$) for the same 
model are also two orders of magnitude smaller than the 
observed excesses. For these estimates, we have
assumed a form $\xi(r,z) = (r/r_0)^{-\gamma}(1+z)^{-3-\epsilon}$
for $r<3 r_0$ and $\xi(r,z)=0$ for $r\geq 3r_0$.
The parameters used for the estimations are 
($r_0\, {\rm [km\, s^{-1}]},\, \gamma$) = 
(1000, 1.8), (2400, 1.8) and (880, 2.2) for the AGN-AGN, 
cluster-cluster (Bahcall 1988) and AGN-cluster (Lilje \& Efstathiou 1988)
correlation functions respectively. At the angular scales of degrees,
the expected correlation is insensitive to the cosmological evolution
of clustering and we have used $\epsilon=0$. 
Since the discrepancies are two orders of magnitude, we also 
expect that the clustering of low X-ray luminosity objects (galaxies, 
low-luminosity AGNs), which have not been considered in the above
calculation, cannot make much contribution, since their 2 -- 10
keV X-ray volume emissivity should be  lower than that of AGNs 
(Miyaji et al. 1994). Thus we expect that the Soltan et al.'s
extended ACF component truely represents a new population of X-ray
structure.  

Here we consider a picture that the component responsible for
the Soltan et al.'s ACF component is also responsible for the
excess zero-lag CCF in T2 and H2. 
We constrain the nature of this component by comparing
the upper-limit to our $\theta=2\fdg5$ CCF with
their ACF and also investigate the contribution of this
component to our zero-lag CCF.
Since the spatial resolution of their ACF measurement is much higher than
our CCF measurement, we have smoothed their ACF with the beam
of our CCF (see Eq. [\ref{eq:etac}]) and compared it with
our CCF. The smallest angle where there is no overlap of the 
beams is at $\theta \sim 2.5\DEG$. At this angle, the smoothed
ACF value is $\sim 1\times 10^{-3}$, while the 2$\sigma$ upper-limit
of our CCF (T2) is $3\times10^{-4}$. Assuming that the detected R6 ACF
is dominated by a single component with a single spectrum, the 
hard-to-soft flux ratio $I_{\rm h}/I_{\rm s}$ of this component is:
\begin{equation}
(I_{\rm h}/I_{\rm s})=(W_{\rm sh}\avrg{I_{\rm h}})/
(W_{\rm ss}\avrg{I_{\rm s}}).
\end{equation}  
The upper limit to this ratio is 1.3 (TOT cts s$^{-1}$)
(R6 cts s$^{-1}$)$^{-1}$. This corresponds
to a Raymond \& Smith plasma with $kT\la 2.5$ keV or
a power-law photon index of $\alpha _{ph}\ga 2.3$ for a galactic
column density of $N_{\rm H}=1.8\times10^{20} {\rm cm^{-2}}$.
 
Assuming that the structure causing the extragalactic 1 keV (R6) ACF 
in Soltan \etal (1995) in the ($0.1\DEG \la \theta \la 3 \DEG$)
comes from a single extended component, we have calculated 
the expected zero-lag R6-A2 CCF ($W_{\rm sh}(0)$). These values
are shown under the models E1 (assuming a $kT=$ 2 keV Raymond-Smith
spectrum) and E2 ($kT=$ 1 keV) (E represents the extended). A heavy metal
abundance of 0.4 (cosmic) has been used in these estimations. 
Since the \ROSAT R6 (1 keV) band is
dominated by the Fe L emission lines,  
this estimate is sensitive to the heavy element abundance
and should be taken as a rough measure. 

\begin{figure}[htbp]
\psfig{file=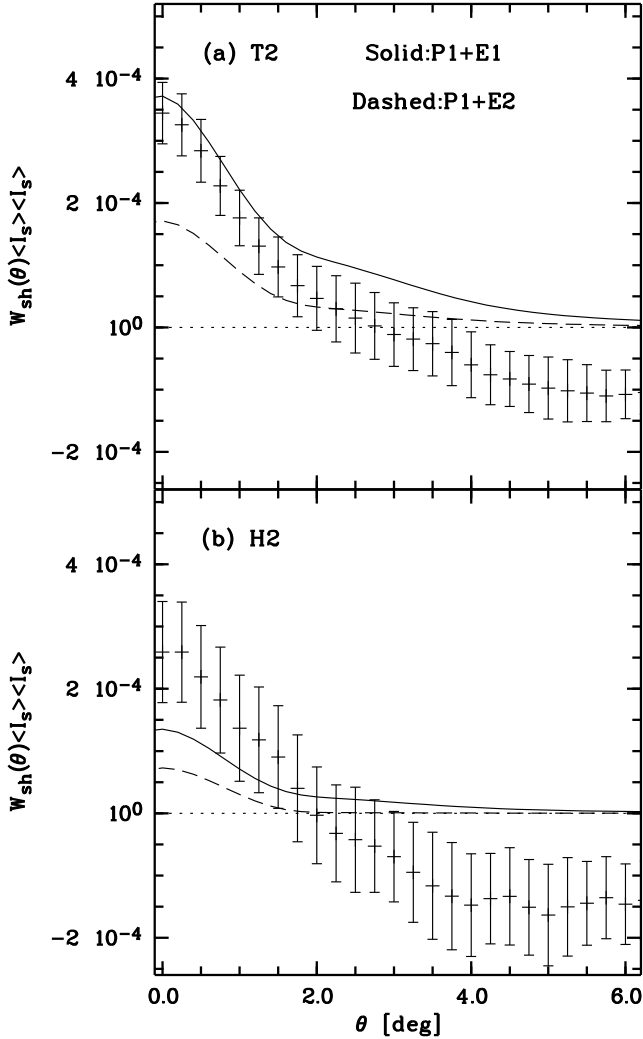,width=\hsize,clip=}    
\caption[]{The observed angular CCFs for the source-removed cases
  are shown with two models as indicated. The 1$\sigma$ error bars
  are from the scatters in the 'rotated' correlations 
  (dotted lines in Fig.\ref{fig:wth}). Note that the errors in 
  data points are not independent but highly correlated}
\label{fig:wth_m}
\end{figure}

Figure \ref{fig:wth_m} shows the comparisons of the observed 
and modeled angular CCF for the source-removed cases. 
The models shown are E1 and E2 in addition to P1 as labeled. 
As we see in Fig. \ref{fig:wth_m},
model (P1+E1) could explain the zero-lag amplitude of the 
CCFs. This means that if the  extended component in the R6 ACF
is a thermal plasma with $kT\sim 2$ keV this would also
explain the excess zero-lag CCF over the model P1 prediction
in our source-removed CCFs within 2$\sigma$. Considering that
the H2 zero-lag correlation still exceeds the model P1+E1, while
statistical significance is poor, a flatter spectrum in the $E\ga 2$ keV
range for this component is preferred. It is possible that the R6 ACF
signal Soltan et al. (1995) found is composed of more than one component.
For example, a composite model with  a small-scale ($\theta \la 2\DEG$) 
hard component and a large-scale ($2\DEG \la \theta \la 5\DEG$) soft 
component would better explain the data.
  
\section{Concluding Discussions}
\label{sec:conc}

As we see from the analysis presented above,  
AGNs and clusters account well for the correlation between 
the \ROSAT bright R6 sources ($\ga 0.08$ R6 cts s$^{-1}$, corresponds
to $2.5\times 10^{-12}\,{\rm erg\,s^{-1}\,cm^{-2}}$) and 
the \HEAO A2 surface brightness. This correlation is dominated by
the Poisson effect from individual clusters of galaxies in the local 
universe. On the other hand, the correlations between the 
residual R6 background and the \HEAO1 A2 surface brightnesses are about a 
factor of 3 -- 4 larger than the Poisson effect expected from 
the population synthesis model by Comastri \etal and clusters
of galaxies based on the known X-ray luminosity functions.
We have considered a few possible components which have 
not been considered in the population synthesis model, i.e.,
AGN-group association, unabsorbed (scattered) soft component 
in type 2 AGNs, and low-luminosity X-ray galaxies. Any of these three are
not likely to explain the observed excess correlation.

Due to the large collimator beam of the \HEAO A2 experiment,
the zero-lag correlation could also be contributed
by the clustering effect or an extended component. 
The extragalactic 1 keV (R6) angular auto-correlation function 
(ACF) by Soltan et al. (1995)
clearly shows an extended structure up to a separation of
several degrees. The origin of this component is not clear at this moment. 
They showed that the effects 
of rich clusters including the cluster-cluster 
clustering and the putative $\sim 10$ Mpc-scale diffuse gas component
in their model to explain the Abell cluster -- 1 keV cross-correlation
function only amounts
to about one third of the R6 ACF signal. They suggested diffuse 
emission from groups of galaxies  
could possibly contribute to the R6 ACF.  
However, statistical properties concerning
poor groups, e.g. luminosity/temperature functions and clustering
properties with rich clusters/AGNs, are scarcely known.   
Comparing the 1 keV ACF with the upper-limit
to our R6 -- A2 (TOT) CCF at $\theta=2.5\DEG$, the minimum angular
separation where beams do not overlap,
we have constrained the spectrum of the extended structure causing
this R6 ACF signal. This corresponds to a Raymond \& Smith plasma of 
$kT \la 2.5$ keV or a power-law photon index of $\alpha_{ph}\ga 2.3$.
Suppose the component responsible for the 1 keV ACF has a spectrum
with an 'equivalent color-temperature' of  $kT \sim$ 2 keV, this
component explains the correlation in excess of the population
synthesis model expectation, \chg{assuming that the extended
R6 ACF structure is composed of a single component. A picture
where there are a harder small-scale component and a softer 
large-scale component would better explain the data.}  
  
 There remains also a possibility that some major change
to the Comastri \etal population synthesis model or an introduction
of a new population is needed, while the extended component has 
a lower temperature (or a steeper spectrum). The current observation 
cannot discriminate between these two, since it is limited by the low spatial
resolution of the \HEAO A2 experiment. A large area survey at 
$E\ga 2$ keV with higher spatial resolution in a future mission such as 
{\it ABRIXAS} (Friedrich et al. 1996)
would enable us to discriminate between these two pictures. With an analysis
similar to that presented here, but with a higher spatial resolution,
we can constrain the spectrum of the extended component and give
another crucial observational constraint to the population synthesis 
models simultaneously. 

\acknowledgements
 TM is supported by a fellowship from the Max-Planck Society. TM 
appreciates the hospitality of the Astrophysikalisches Institut Potsdam 
during his visit. The authors thank Keith Jahoda for his help
in handling the \HEAO A2 data and Elihu Boldt for encouragements. 
This work has made use of data from two X-ray satellites, 
\ROSAT and \HEAO, in orbit about
a decade apart from each other. The authors appreciate the effort 
of all people which lead to the success of these two missions.
Thanks are also due to the referee, Luigi Danese, for useful 
comments and suggestions. 


\begin{thebibliography}{}
%
  \bibitem{} Allen J., Jahoda K.,\&  Whitlock, L. 1994, Legacy 5, 27  

  \bibitem{} Awaki H. Koyama K., Inoue H., \& Halpern J.P. 
       1991, PASJ 43, 195

  \bibitem{} Bahcall N.A. 1988, ARA\&A 26, 631   

  \bibitem{} Barcons X., Branduardi-Raymont G., Warwick, R.S., Fabian, A.C.,
        Mason K. O., McHardy I., \&  Rowan-Robinson M., 1994, 
        MNRAS, 268, 833

  \bibitem{} Barcons X., Franceschini A. Danese L., De Zotti G. \&
           Miyaji T. 1995 ApJ, 455, 480

  \bibitem{} Boyle B.J., Griffiths R.E., Shanks T., Stewart G.C.
         \& Georgantopoulos I. 1993, MNRAS 260, 49

  \bibitem{} Boyle B.J., Shanks T., Georgantopoulos I., 
          Stewart, G.C., \& Griffiths, R.E. 1994, MNRAS 271, 639

  \bibitem{} Carrera F.J. \etal 1991, MNRAS 249, 698

  \bibitem{} Comastri A., Setti G., Zamorani G., \& Hasinger G.
          1995, A\&A 296, 1
  
  \bibitem{} Danese L., Toffolatti L., Franceschini A., 
	Mart\'in-Mirones J.M., \& De Zotti G. 1993, ApJ, 412, 56  

  \bibitem{} David L. P., Slyz A., Jones C., Forman W., 
	   Vrtilek S. D., \& Arnaud, K. A. 1993, ApJ 412, 479 

  \bibitem{} Edge A. C., Stewart G. C., Fabian A. C., \& 
	   Arnaud K. A. 1990, MNRAS 245, 559

  \bibitem{} Friedrich P. et al. 1996, in R\"ontgenstrahlung 
       from the Universe, eds. Zimmermann H.U. \& Tr\"umper J., 
        MPE report, in press

  \bibitem{} Gioia I.M., Maccacaro T., Schild R.E., Wolter A., 
        Stocke J.T., Morris S.L. \& Henry J.P. 1990, ApJS, 72, 567

  \bibitem{} Hasinger G., Burg R. Giacconi R. Hartner G., 
           Schmidt M., Tr\"umper J., \& Zamorani G. 1993, A\&A 275,1
  
  \bibitem{} Henry J.P. \etal 1992, ApJ 386, 408

  \bibitem{} Henry J.P. \etal 1995, ApJ 449, 422

  \bibitem{} Jahoda K. 1993, Adv. Space Res. 13, No.12, 231

  \bibitem{} Jahoda K., Lahav O., Mushotzky R., Boldt E. 1991, ApJ 378, 
		L37 

  \bibitem{} ---. 1992, ApJ 399, L107 (Erratum) 

  \bibitem{} Lahav O. \etal 1993, Nature 364, 693
  
  \bibitem{} Lilje P.J. \& Efstathiou G. 1988, MNRAS 231, 635

  \bibitem{} Madau P., Ghisellini G. \& Fabian A.C. 1994, MNRAS 270, L17


  \bibitem{} Miyaji T. 1994, PhD Thesis, University of Maryland

  \bibitem{} Miyaji T., Lahav O., Jahoda K., \& Boldt E., 1994,
          ApJ 434, 424

  \bibitem{} Mulchaey J.S., Mushotzky, R.F., \& Weaver, K.A. 1992
           ApJ 390, L69

  \bibitem{} Pfeffermann, E. \etal 1986, Proc SPIE 733, 519

  \bibitem{} Piccinotti G.  \etal 1982, ApJ 253, 485

  \bibitem{} Roche N., Shanks, T., Geaorgantopuolos, I.,
         Boyle B.J., \& Griffiths, R.E. 1995, MNRAS 273, L15

  \bibitem{} Rothschild R.E. \etal 1979, Sp. Sci. Instr., 4, 269

  \bibitem{} Schartel N, Schmidt M., Fink H. H., Hasinger G., \&
          Tr\"umper J. 1995, preprint

  \bibitem{} Shafer R. A. 1983 PhD Thesis, University of Maryland 

  \bibitem{} Soltan A. \& Hasinger G. 1994, A\&A 288, 77 

  \bibitem{} Soltan A., Hasinger G., Egger R., Snowden S.,
         \& Tr\"umper J. 1995, A\&A, in press 

  \bibitem{} Snowden S.L. \& Schmitt, J.H.M.M. 1990, Ap\&SS, 171, 207

  \bibitem{} Snowden S.L., McCammon D., Burrows D.N., Mendenhall J.A.
         1994, ApJ 424, 714

  \bibitem{} Snowden S.L. \etal 1995, ApJ in press

  \bibitem{} Stocke J.T., Morris S.L., Gioia I.M., Maccacaro T.,
          Schild R., Wolter A., Fleming T.A., \& Henry, J.P. 1991, ApJS
          76, 813

 \bibitem{} Tr\"umper J. Adv. Space Res. 2, No.4, 241

 \bibitem{} Voges W. 1992, in Proceedings of Satellite Symposium 3, 
        Space Science with particular emphasis on High-Energy Astrophysics, 
        ed. T.D. Guyenne \& J.J. Hunt (noordwijk:ESA Publication Division), 9

 \bibitem{} Voges W. 1995,  private communication

 \bibitem{} Yaqoob T., Serlemitsos P.J., Ptak, A., Mushotzky R., 
         Kunieda H., \& Terashima Y. 1995, ApJ, 455, 508 

 \bibitem{} Zimmermann H.U., Becker W., Belloni T., D\"obereiner S.,
         Izzo, C., Kahabka P. \& Schwentker O. 1994, EXSAS User's Guide,
         MPE Report 257, Garching
      
%
\end{thebibliography}
\end{document}